\documentclass[pre,twocolumn,superscriptaddress]{revtex4-1}

\usepackage{amsmath}
\usepackage{graphicx}
\usepackage{color}
\usepackage{hyperref}
\begin{document}

\title{Social Diffusion and Global Drift on Networks}

\author{Hiroki Sayama}
\affiliation{Collective Dynamics of Complex Systems Research Group, Binghamton University, Binghamton, New York 13902, USA}
\affiliation{Center for Complex Network Research and Department of Physics, Northeastern University, Boston, Massachusetts 02115, USA}

\author{Roberta Sinatra}
\affiliation{Center for Complex Network Research and Department of Physics, Northeastern University, Boston, Massachusetts 02115, USA}
\affiliation{Center for Cancer Systems Biology, Dana-Farber Cancer Institute, Boston, MA 02115, USA}

\date{\today}

\begin{abstract}
We study a mathematical model of social diffusion on a symmetric
weighted network where individual nodes' states gradually assimilate
to local social norms made by their neighbors' average states. Unlike
physical diffusion, this process is not state conservational and thus
the global state of the network (i.e., sum of node states) will
drift. The asymptotic average node state will be the average of
initial node states weighted by their strengths. Here we show that,
while the global state is not conserved in this process, the inner
product of strength and state vectors is conserved instead, and perfect
positive correlation between node states and local averages of their
self/neighbor strength ratios always results in upward (or at least
neutral) global drift. We also show that the strength assortativity
negatively affects the speed of homogenization. Based on these
findings, we propose an adaptive link weight adjustment method to
achieve the highest upward global drift by increasing the strength-state
correlation. The effectiveness of the method was confirmed through
numerical simulations and implications for real-world social applications
are discussed.
\end{abstract}

\maketitle

\section{Introduction}

Social contagion \cite{christakis2013social} has been studied in
various contexts. Many instances of social contagion can be modeled as
an infection process where a specific state (adoption of product, fad,
knowledge, behavior, political opinion, etc.)  spreads from individual
to individual through links between them
\cite{van2004social,dodds2004universal,hill2010infectious,bond201261}. In
the meantime, other forms of social contagion may be better understood
as a diffusion process where the state of an individual tends to
assimilate gradually to the social norm (i.e., local average state)
within his/her neighborhood
\cite{christakis2007spread,fowler2008dynamic,coronges2011social,coviello2014detecting}.

Unlike infection scenarios where influence is nonlinear,
unidirectional, fast, and potentially disruptive, social diffusion is
linear, bidirectional, gradual, and converging. The distance between
an individual's state and his/her neighbors' average state always
decreases, and thus a homogeneous global state is guaranteed to be the
connected network's stable equilibrium state in the long run \cite{castellano2009statistical}. 

Here, we focus on an unrecognized characteristic of social
diffusion, i.e., non-trivial drift it can cause to the network's
global state. Although somewhat counterintuitive, such global drift is
indeed possible because, unlike physical diffusion processes, social
diffusion processes are {\em not} state conservational. In what
follows, we study a simple mathematical model of social diffusion to
understand the mechanisms of this process and obtain both asymptotic
and instantaneous behaviors of the global state of the network. We
also show how strength assortativity influences the speed of
homogenization. Then we propose an adaptive network
\cite{gross2009adaptive} method of preferential link weight adjustment
to achieve the highest upward global drift within a given time
period. The relevance of social diffusion to individual and collective
improvement is discussed, with a particular emphasis on educational
applications.

\section{Mathematical Model of Social Diffusion}

Let us begin with the conventional physical diffusion equation on a
simple symmetric network,
\begin{equation}
\frac{ds}{dt} = - c L s , \label{physical}
\end{equation}
where $s$ is the node state vector of the network, $c$ the diffusion
constant, and $L$ the Laplacian matrix of the network. The Laplacian matrix of a network is defined as $L=D-A$, where $A$ is the adjacency matrix of the network and $D$ is a diagonal matrix of the nodes' degrees. If the network
is connected, the coefficient matrix $(- c L)$ has one and only one
dominant eigenvalue, $\lambda _0=0$, whose corresponding eigenvector is the
homogeneity vector $h = (1 \; 1 \; 1 \; \ldots \; 1)^T$. Therefore,
the solutions of this equation always converge to a homogeneous state
regardless of initial conditions if the network is connected. It is
easy to show that this process is state conservational, i.e., the sum
(or, equivalently, average) of node states, $h^Ts$, will not change
over time:
\begin{equation}
\frac{d(h^T s)}{dt} = - c h^T L s = - c (Lh)^T s = 0
\end{equation}
This indicates that no drift of the global state is possible in
physical diffusion processes on a simple symmetric network.

Social diffusion processes can be modeled differently. In this
scenario, a state of an individual node may represent any quantitative
property of the individual that is subject to social influence from
his/her peers, such as personal preference, feeling, cooperativeness,
etc. We assume that each individual gradually assimilates his/her
state to the social norm around him/her. The dynamical equation of
social diffusion can be written individually as
\begin{equation}
\frac{ds_i}{dt} = c \left( \langle s_j \rangle^i_j - s_i \right) , \label{socialdiffusion}
\end{equation}
where $s_i$ is the state of individual $i$, and $\langle x_j \rangle^i_j$
is a local weighted average of $x_j$ around individual $i$, i.e.,
\begin{equation}
\langle x_j \rangle^i_j = \frac{\sum_j a_{ij} x_j}{\sum_j a_{ij}} , \label{socialnorm}
\end{equation}
with $a_{ij} \ge 0$ being the connection weight from individual $j$ to
individual $i$. In Eq.~(\ref{socialdiffusion}) $x_j = s_j$, but this
weighted averaging will also be used for other local quantities later
in this paper. We note that Eq.~(\ref{socialnorm}) carries a dummy
variable $j$ {\em outside} the brackets. While not customary, such explicit
specification of dummy variable $j$ is necessary in the present study,
in order to disambiguate which index the averaging is conducted over
when calculating nested local averages (to appear later).

$\langle s_j \rangle^i_j$ in
Eq.~(\ref{socialdiffusion}) represents the social norm for individual
$i$. Discrete-time versions of similar models are also used to
describe peer effects in opinion dynamics in sociology
\cite{friedkin1990social}, distributed consensus formation (called
``agreement algorithms'') in control and systems engineering
\cite{bertsekas1989parallel,blondel2005convergence,lambiotte2011flow},
and for statistical modeling of empirical social media data more
recently \cite{coviello2014detecting}.

In this paper, we focus on symmetric interactions between individuals
(i.e., $a_{ij}=a_{ji}$). Such symmetry is a reasonable assumption as a
model of various real-world collaborative or contact relationships,
e.g., networks of coworkers in an organization, networks of students
in a school, and networks of residents in a community, to which
symmetric social interaction can be relevant. Self-loops are allowed
in our model, i.e., $a_{ii}$ may be non-zero. Such self-loops
represent self-confidence of individuals when calculating their local
social norms. The sum of weights of all links attached to a node $i$, $k_i = \sum_j a_{ij} = \sum_j a_{ji}$, is called the {\em strength} of node $i$ \cite{barrat2004architecture}, which would correspond to a node degree for unweighted networks. This social diffusion model requires $k_i \ne 0$ $\forall i$;
otherwise Eq.~(\ref{socialnorm}) would be indeterminate and there
would be no meaningful dynamics to be described for the individual.

Eq.~(\ref{socialdiffusion}) can be rewritten at a collective level as
\begin{equation}
\frac{ds}{dt} = c (D^{-1}A - I) s , \label{socialdiffusion2}
\end{equation}
where $A = (a_{ij})$ and $D^{-1}$ is a square matrix whose $i$-th
diagonal component is $k_i^{-1}$ while non-diagonal components are all
zero. If the network does not have link weights or self-loops (i.e.,
$a_{ij} \in \{0,1\}$, $a_{ii} = 0$), $D^{-1}$ and $A$ are conventional
inverse degree and adjacency matrices, respectively, and thus the
matrix $D^{-1}A - I = - D^{-1}L$ is a special case of the coupling
matrix discussed in \cite{motter2005}, with scaling exponent 1 and a
negative sign added.

\section{Asymptotic Behaviors}

Eq.~(\ref{socialdiffusion2}) is essentially the same as
Eq.~(\ref{physical}) if the network is regular without link weights or
self-loops (i.e., $a_{ij} \in \{0,1\}$, $a_{ii} = 0$, $k_i = k$
$\forall i$). Even if not regular, it is still a simple matrix
differential equation, for which a general solution is always
available. If the network is connected, the coefficient matrix $c
(D^{-1}A - I)$ has one and only one dominant eigenvalue 0 with
corresponding eigenvector $h$, hence the state of the network will
always converge to a homogeneous equilibrium state, just like in the
physical diffusion equation. Hence the asymptotic state can be written
as
\begin{equation}
s_\infty = \lim_{t \to \infty} s(t) = \langle s_\infty \rangle h , \label{asymptoticstate}
\end{equation}
where $\langle s_\infty \rangle$ is the average node state in
$s_\infty$.

It is known that a discrete time version of
Eq.~(\ref{socialdiffusion2}) will converge to a weighted average of
initial node states with their strengths used as weights, if the network
is connected and non-bipartite \cite{lambiotte2011flow}.
The same conclusion was also reported for the ensemble average of
the voter model with a node-update scheme \cite{suchecki2005conservation}.
It can be
easily shown that our continuous-time
model has the same asymptotic state, as follows.

First, we show that the inner product of strength and state vectors,
$g^Ts$ with $g = (k_1 \; k_2 \; \ldots \; k_n)^T$ (where $n$ is the
number of individuals), is always conserved instead of node states
during social diffusion on a symmetric network, because
\begin{equation}
\frac{d(g^Ts)}{dt} = cg^TD^{-1}As - cg^Ts = ch^TAs -cg^Ts = 0 . \label{gTs-constant}
\end{equation}
This holds for any symmetric networks regardless of their topologies
and link weights. We call $g^Ts$ a {\em strength-state product}
hereafter.

This conservation law was already known for the ensemble average of
the node-update voter model \cite{suchecki2005conservation}, and the
existence of similar conservation laws was also shown for more
generalized voter-like models with directed links
\cite{serrano2009conservation}. Our result above provides an example
of the same conservation law realized in a different model setting,
i.e., continuous-time/state social diffusion models on weighted,
undirected networks.

Calculating an inner product of each side of
Eq.~(\ref{asymptoticstate}) with $g$ makes
\begin{equation}
g^T s_\infty = \langle s_\infty \rangle g^T h .
\end{equation}
The conservation of the strength-state product
(Eq.~(\ref{gTs-constant})) allows us to replace the left hand side,
resulting in
\begin{eqnarray}
g^T s_0 &=& \langle s_\infty \rangle g^T h , \\
\langle s_\infty \rangle &=& g^T s_0 / g^T h \\
&=& (k_1/K \; k_2/K \; \ldots \; k_n/K)^T s_0 ,
\end{eqnarray}
where $K = g^T h = \sum_i k_i$. This shows that the asymptotic average
node state is a weighted average of initial node states where their
strengths are used as weights.

In discrete-time distributed consensus formation models, the network
needs to be non-bipartite (in addition to being connected) for global
homogenization to be reached \cite{lambiotte2011flow}, but the model
discussed here does not require non-bipartiteness because time is
continuous.

The overall net gain or loss of the global state that can be attained
asymptotically by social diffusion is calculated as
\begin{eqnarray}
\Delta_\infty &=& h^T s_\infty - h^T s_0 \\
&=& h^T \left( \frac{g^T s_0}{g^T h} h \right) - h^Ts_0 \\
&=& (\hat{g} - h)^Ts_0 , \label{netgain}
\end{eqnarray}
where $\hat{g} = g / \langle k \rangle$ with $\langle k \rangle = g^T
h / n$ (i.e., average strength). This means that the asymptotic net gain
or loss will be determined by the correlation between the initial
state vector $s_0$ and another vector $(\hat{g} - h)$ that is
determined by the node strengths.

\section{Instantaneous Behaviors}

Next, we study the instantaneous direction of drift of the global
state (sum of node states, $h^T s$), which can be
written as
\begin{eqnarray}
\frac{d(h^T s)}{dt} &=& ch^T(D^{-1}A-I)s \\
&=& c (w^T s) - c (h^T s) , \label{direction-hTs}
\end{eqnarray}
where $w = A D^{-1} h$. $w$ can be further detailed as
\begin{equation}
w =
\left(\begin{array}{c}
\sum_j a_{1j} k_j^{-1} \\
\sum_j a_{2j} k_j^{-1} \\
\vdots \\
\sum_j a_{nj} k_j^{-1} \\
\end{array}\right) \\
=
\left(\begin{array}{c}
\langle k_1 / k_j \rangle^1_j \\
\langle k_2 / k_j \rangle^2_j \\
\vdots \\
\langle k_n / k_j \rangle^n_j \\
\end{array}\right) . \label{wcontents}
\end{equation}
Each component of $w$ is the local average of self/neighbor strength
ratios, $k_i/k_j$, which tends to be greater than 1 if the individual
has more connection weights than its neighbors, or less than 1
otherwise (but always strictly positive). In this regard, $w$
characterizes the local strength differences for all individuals in
society. If $(w - h)^Ts > 0$, the global state will drift upward due to
social diffusion. Also note that $w^Th = n$, because
\begin{eqnarray}
w^Th &=& \sum_i \sum_j a_{ij} k_j^{-1} \\
&=& \sum_j \left( k_j^{-1} \sum_i a_{ij} \right) =
\sum_j k_j^{-1} k_j = n .
\end{eqnarray}

We show $w \approx \hat{g}$ for networks with neutral strength
assortativity (called {\em non-assortative networks} hereafter). 
Let $P(k' | k)$ be the
conditional probability density for a link originating from a $k$-degree node to
reach a $k'$-degree node. Then, each
component of $w$ in Eq.~(\ref{wcontents}) is approximated as
\begin{equation}
\langle k_i / k_j \rangle^i_j \approx k_i \int_{k'} k'^{-1} P(k'|k_i) dk' . \label{knn-1}
\end{equation}
For non-assortative networks, $P(k'|k_i)$ does not depend on $k_i$ \cite{barabasibook}:
\begin{equation}
P(k'|k_i) = k' P(k') / \int_{k'} k'P(k')dk' = k' P(k') / \langle k \rangle \label{pkkneutral}
\end{equation}
Applying Eq.~(\ref{pkkneutral}) to Eq.~(\ref{knn-1}) makes
\begin{equation}
\langle k_i / k_j \rangle^i_j \approx k_i \int_{k'} k'^{-1} k' P(k') dk' /
\langle k \rangle = k_i / \langle k \rangle . \label{kikjapprox}
\end{equation}
With this,
Eq.~(\ref{wcontents}) 
is
approximated as
\begin{equation}
w \approx g / \langle k \rangle = \hat{g}. \label{wcontents2}
\end{equation}
Using $\hat{g}$, we obtain the following approximated dynamical equation
for non-assortative networks, which is quite similar to Eq.~(\ref{netgain}):
\begin{equation}
\frac{d(h^T s)}{dt} \approx c (\hat{g}^T s) - c (h^T s) = c(\hat{g} - h)^Ts 
\end{equation}

\begin{figure}
\includegraphics[width=0.5\textwidth]{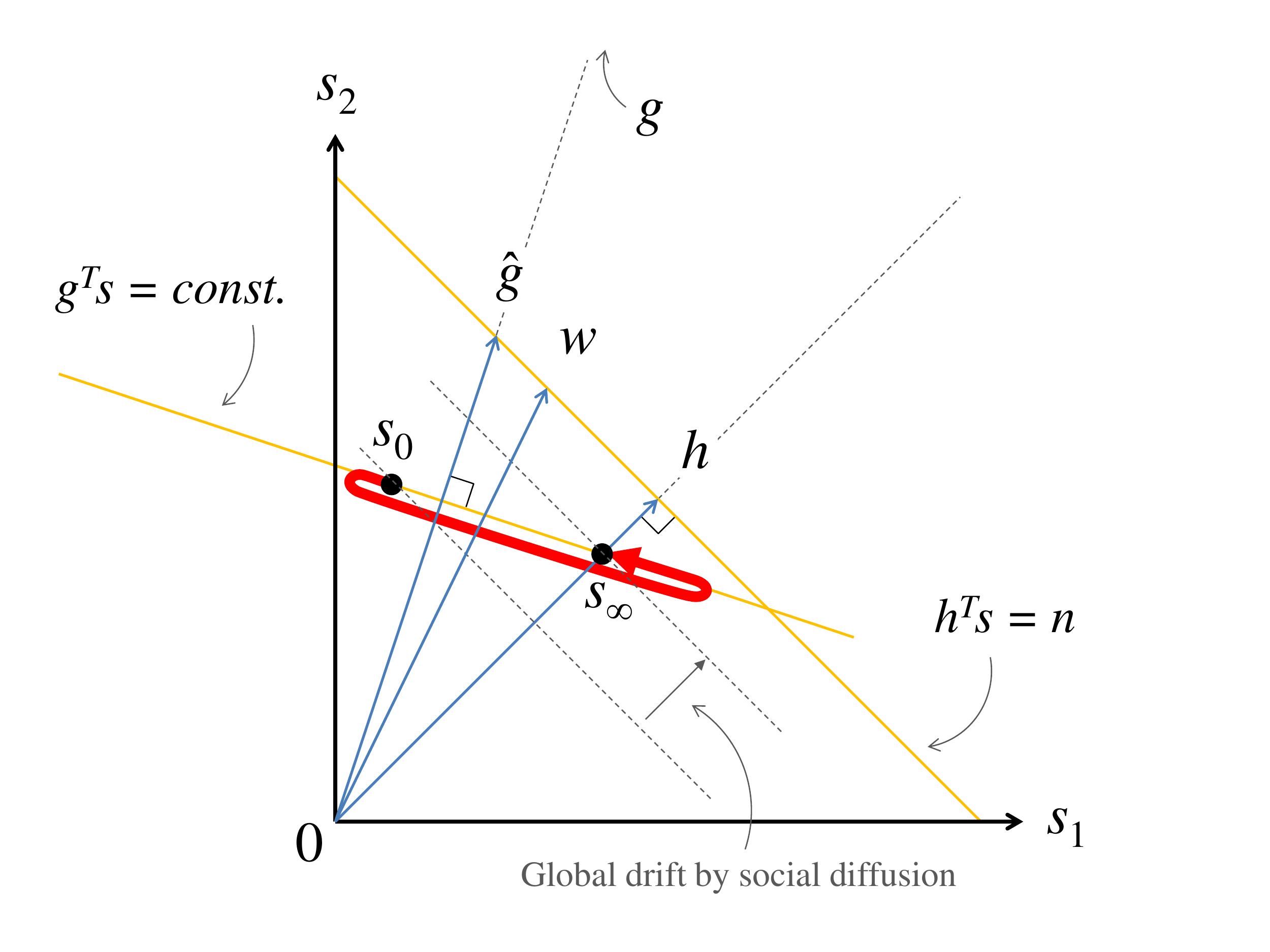}
\caption{(Color online) Visual illustration of the relationships between several
  vectors discussed ($h$, $w$, $g$, $\hat{g}$, $s_0$,
  $s_\infty$). This diagram is drawn in two dimensions for simplicity,
  but the actual vectors exist in an $n$-dimensional state space. Note
  that $h^Th = w^Th = \hat{g}^Th=n$. If the network is non-assortative, then $w \approx \hat{g}$.}
\label{vectors}
\end{figure}

Furthermore, we prove that perfect positive correlation between node
states ($s$) and local averages of their self/neighbor strength ratios
($w$) always results in upward (or at least neutral) global drift. If
those two vectors are in perfect positive correlation, $s = \alpha w$
with positive constant $\alpha$. Then Eq.~(\ref{direction-hTs})
becomes
\begin{equation}
\frac{d(h^T s)}{dt} = \alpha c (w - h)^T w . \label{proportionalcase}
\end{equation}
Here, $w$ and $w-h$ are the hypotenuse and adjacent of a right
triangle in an $n$-dimensional state space, respectively, because
\begin{equation}
(w-h)^T h = w^T h - h^T h = n - n = 0 ,
\end{equation}
which shows $(w-h) \perp h$. Therefore, the angle between $(w-h)$ and
$w$ cannot be greater than $\pi/2$, which guarantees in
Eq.~(\ref{proportionalcase}) that $d(h^Ts)/dt$ is always
non-negative. Figure \ref{vectors} visually illustrates the
relationships between the vectors discussed above.

\section{Approximation as Low-Dimensional Linear Dynamical System and Analysis of Homogenization Speed}

Here we analyze the speed of homogenization caused by social diffusion
on networks by approximating the whole dynamics in a low-dimensional
linear dynamical system about $h^Ts$ and $w^Ts$. The dynamics of
$w^Ts$ is given by
\begin{eqnarray}
\frac{d(w^T s)}{dt} &=& c w^T D^{-1} (A-D) s \\ &=& c (u^T s) - c (w^Ts) , \label{direction-wTs}
\end{eqnarray}
where $u=AD^{-1}w$, which is further detailed as
\begin{equation}
u = A
\left(\begin{array}{c}
\langle k_j^{-1} \rangle^1_j \\
\langle k_j^{-1} \rangle^2_j \\
\vdots \\
\langle k_j^{-1} \rangle^n_j \\
\end{array}\right)
=
\left(\begin{array}{c}
\langle \langle k_1 / k_j \rangle^i_j \rangle^1_i \\
\langle \langle k_2 / k_j \rangle^i_j \rangle^2_i \\
\vdots \\
\langle \langle k_n / k_j \rangle^i_j \rangle^n_i \\
\end{array}\right) . \label{uvector}
\end{equation}
If we can approximate $u$ using $h$ and/or $w$, the system closes and
a simple eigenvalue analysis reveals the speed of homogenization.

For non-assortative networks, the approximation in Eq.~(\ref{kikjapprox})
gives
\begin{equation}
u \approx g / \langle k \rangle = \hat{g} . \label{uvector-neutral}
\end{equation}
Combining this with Eq.~(\ref{wcontents2}), we obtain
\begin{equation}
\frac{d(w^T s)}{dt} \approx c(\hat{g}^Ts) - c(\hat{g}^Ts) = 0 ,
\end{equation}
which is consistent with our analyses derived earlier. Combining this
result with Eq.~(\ref{direction-hTs}) makes the following
two-dimensional linear dynamical system:
\begin{equation}
\frac{d}{dt}\begin{pmatrix}h^Ts\\w^Ts\end{pmatrix} =
  c \left(\begin{array}{rr} -1 & 1 \\ 0&0 \end{array}\right)
\begin{pmatrix}h^Ts \\ w^Ts\end{pmatrix}
\end{equation}
The coefficient matrix above has eigenvalues $0$ and $-c$ with
corresponding eigenvectors $(1 \; 1)^T$ and $(1 \; 0)^T$,
respectively. The second eigenvalue ($-c$) represents the baseline
speed of homogenization on non-assortative networks.

This low-dimensional dynamical systems approach can be extended to
networks with negative or positive strength assortativity. In so doing,
we adopt a scaling model \cite{gallos2008scaling, barabasibook} to approximate the strength
of a neighbor of node $i$ by a scaling function of $k_i$,
\begin{equation}
k_j \approx b k_i^\mu , \label{knnapprox}
\end{equation}
where $\mu$ is the correlation exponent and $b$ is a positive
constant. Applying this approximation to Eq.~(\ref{uvector}) gives
\begin{eqnarray}
u &\approx&
\left(\begin{array}{c}
k_1 / (b (b k_1^\mu)^\mu)\\
k_2 / (b (b k_2^\mu)^\mu)\\
\vdots \\
k_n / (b (b k_n^\mu)^\mu)\\
\end{array}\right)%\\
= b^{-1-\mu}
\left(\begin{array}{c}
k_1^{1-\mu^2}\\
k_2^{1-\mu^2}\\
\vdots \\
k_n^{1-\mu^2}\\
\end{array}\right) . \label{uvector-approx}
\end{eqnarray}
This approximation indicates that, if $\mu \to 0$, $u \to g/b$, which
agrees with Eq.~(\ref{uvector-neutral}) for non-assortative networks. For
simplicity, we limit our analysis to two unrealistic yet illustrative
cases with $\mu = \pm 1$, because this setting makes $1-\mu^2 = 0$
with which the dynamics can still be written using $h$ and $w$ only.

For strongly disassortative networks ($\mu = -1$),
Eq.~(\ref{uvector-approx}) becomes $u \approx h$. The resulting linear
dynamical system is
\begin{equation}
\frac{d}{dt}\begin{pmatrix}h^Ts \\ w^Ts\end{pmatrix} =
  c \left(\begin{array}{rr} -1 & 1 \\ 1& -1\end{array}\right)
\begin{pmatrix}h^Ts \\ w^Ts\end{pmatrix} ,
\end{equation}
whose coefficient matrix has eigenvalues $0$ and $-2c$ with
corresponding eigenvectors $(1 \; 1)^T$ and $(-1 \; 1)^T$,
respectively. The second eigenvalue ($-2c$) is smaller than that of
non-assortative networks, which shows that the homogenization takes place
faster on disassortative networks.

Finally, for strongly assortative networks ($\mu = 1$), applying the
approximation Eq.~(\ref{knnapprox}) with $\mu=1$ to
Eq.~(\ref{wcontents}) gives $w \approx h/b$, and also
Eq.~(\ref{uvector-approx}) gives $u \approx h/b^2$. This means that
the system is essentially collapsed into the following one-dimensional
linear dynamical system:
\begin{eqnarray}
\frac{d(h^Ts)}{dt} &=& c((h/b)^Ts) - c (h^Ts) \\
&=& -c \frac{b - 1}{b} (h^Ts) \label{direction-assortative}
\end{eqnarray}
For any positive $b$ and $c$, always
\begin{equation}
-c \frac{b-1}{b} > - c .
\end{equation}
This shows that the homogenization takes place slower on assortative
networks than on non-assortative networks. Moreover, if the network is truly
strongly assortative, $b$ must be close to 1 by definition. This
brings the coefficient in Eq.~(\ref{direction-assortative}) close to
0. Therefore the homogenization process on a strongly assortative
network must be extremely slow. This can also be understood
intuitively; extreme assortativity would result in having links only
between nodes of equal strength, making the network disconnected and
thus stopping the homogenization process.

These analytical results collectively illustrate that, as the strength
assortativity increases, the speed of homogenization goes down. This
finding is consistent with the negative effect of degree assortativity
on the spectral gap of the Laplacian matrix reported in
\cite{van2010influence} and on entropy measures of biased random walks reported in \cite{sinatra2011maximal}, and also similar to the enhancement of
synchronizability of networked nonlinear oscillators by negative
degree assortativity
\cite{sorrentino2006synchronization,di2007effects, arenas2008synchronization}.

\begin{figure*}
\centering
\begin{tabular}{ccc}
\includegraphics[width=0.49\textwidth]{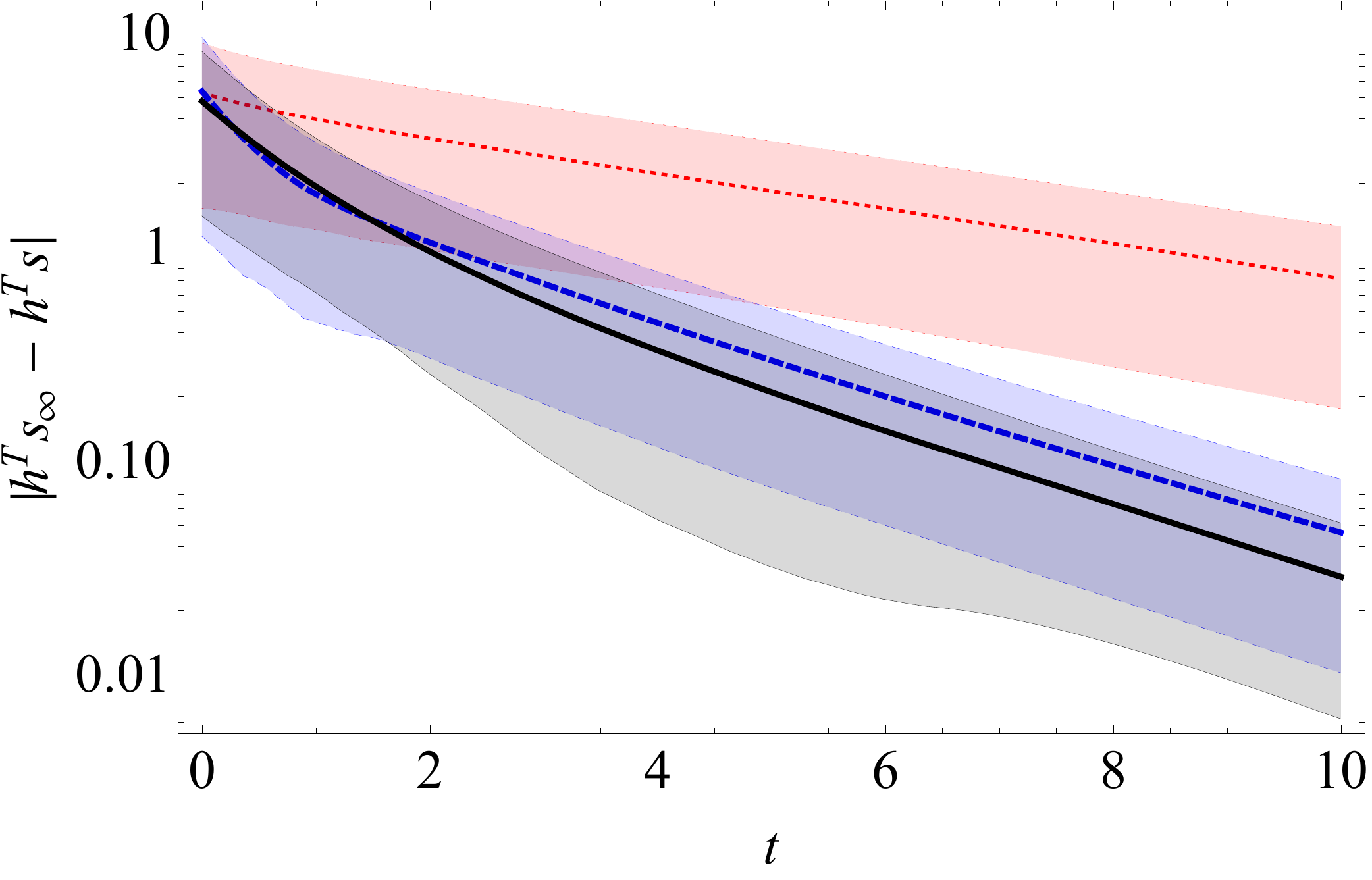} & ~ &
\includegraphics[width=0.49\textwidth]{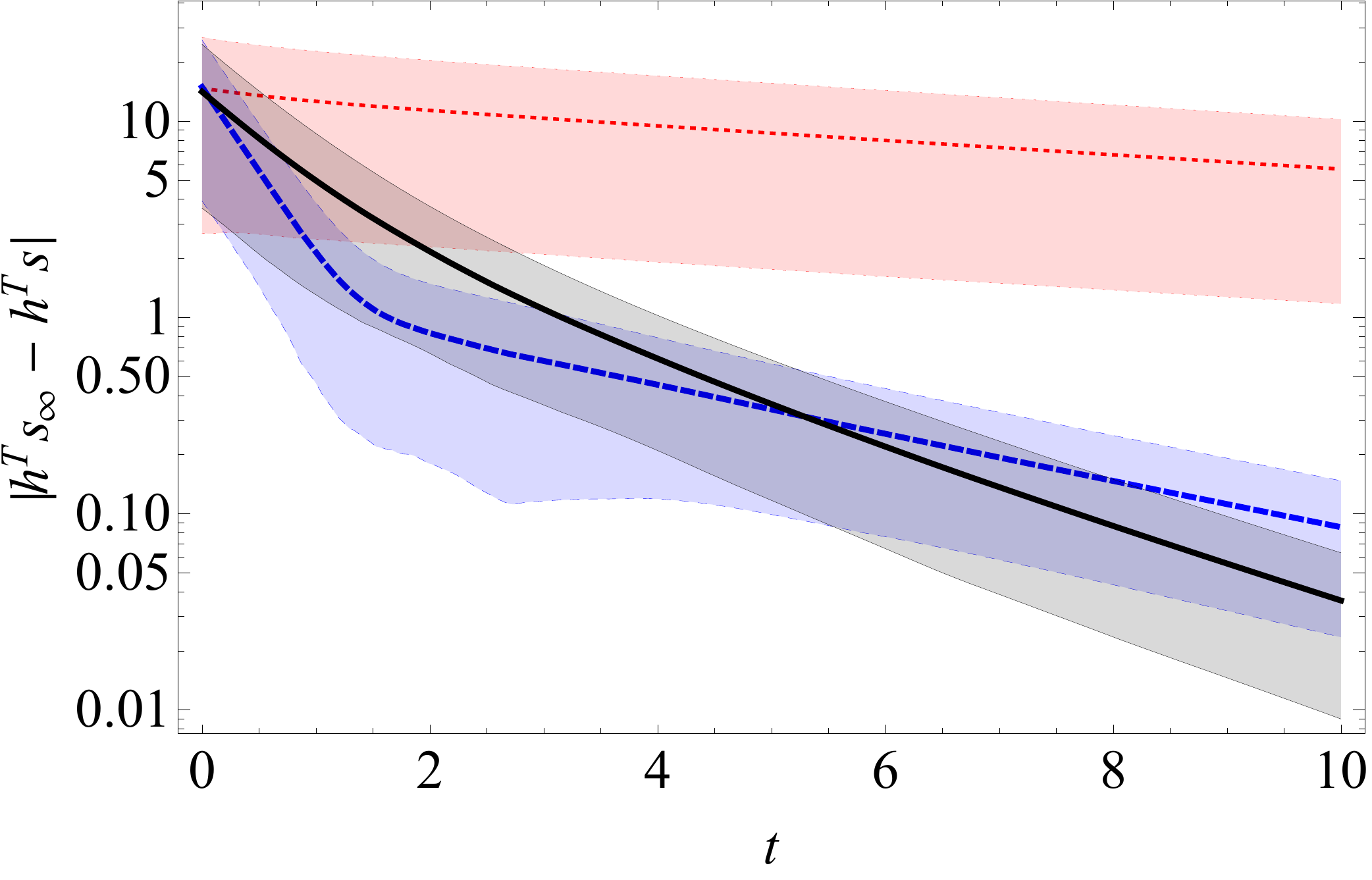} \\
(a) random & ~ & (b) scale-free
\end{tabular}
~\\
\textcolor{black}{\rule{.5in}{1.4pt}} : non-assortative networks
\hspace{1cm}
\textcolor{blue}{\rule{3pt}{1.4pt}}%
\textcolor{white}{\rule{0.5pt}{1.4pt}}%
\textcolor{blue}{\rule{3pt}{1.4pt}}%
\textcolor{white}{\rule{0.5pt}{1.4pt}}%
\textcolor{blue}{\rule{3pt}{1.4pt}}%
\textcolor{white}{\rule{0.5pt}{1.4pt}}%
\textcolor{blue}{\rule{3pt}{1.4pt}}%
\textcolor{white}{\rule{0.5pt}{1.4pt}}%
\textcolor{blue}{\rule{3pt}{1.4pt}}%
\textcolor{white}{\rule{0.5pt}{1.4pt}}%
\textcolor{blue}{\rule{3pt}{1.4pt}}%
\textcolor{white}{\rule{0.5pt}{1.4pt}}%
\textcolor{blue}{\rule{3pt}{1.4pt}}%
\textcolor{white}{\rule{0.5pt}{1.4pt}}%
\textcolor{blue}{\rule{3pt}{1.4pt}}%
\textcolor{white}{\rule{0.5pt}{1.4pt}}%
\textcolor{blue}{\rule{3pt}{1.4pt}}%
\textcolor{white}{\rule{0.5pt}{1.4pt}}%
\textcolor{blue}{\rule{3pt}{1.4pt}}%
\textcolor{white}{\rule{0.5pt}{1.4pt}}%
\textcolor{blue}{\rule{3pt}{1.4pt}}%
\textcolor{white}{\rule{0.5pt}{1.4pt}}%
: disassortative networks
\hspace{1cm}
\textcolor{red}{\rule{1.5pt}{0.9pt}}%
\textcolor{white}{\rule{1.1pt}{0.9pt}}%
\textcolor{red}{\rule{1.5pt}{0.9pt}}%
\textcolor{white}{\rule{1.1pt}{0.9pt}}%
\textcolor{red}{\rule{1.5pt}{0.9pt}}%
\textcolor{white}{\rule{1.1pt}{0.9pt}}%
\textcolor{red}{\rule{1.5pt}{0.9pt}}%
\textcolor{white}{\rule{1.1pt}{0.9pt}}%
\textcolor{red}{\rule{1.5pt}{0.9pt}}%
\textcolor{white}{\rule{1.1pt}{0.9pt}}%
\textcolor{red}{\rule{1.5pt}{0.9pt}}%
\textcolor{white}{\rule{1.1pt}{0.9pt}}%
\textcolor{red}{\rule{1.5pt}{0.9pt}}%
\textcolor{white}{\rule{1.1pt}{0.9pt}}%
\textcolor{red}{\rule{1.5pt}{0.9pt}}%
\textcolor{white}{\rule{1.1pt}{0.9pt}}%
\textcolor{red}{\rule{1.5pt}{0.9pt}}%
\textcolor{white}{\rule{1.1pt}{0.9pt}}%
\textcolor{red}{\rule{1.5pt}{0.9pt}}%
\textcolor{white}{\rule{1.1pt}{0.9pt}}%
\textcolor{red}{\rule{1.5pt}{0.9pt}}%
\textcolor{white}{\rule{1.1pt}{0.9pt}}%
\textcolor{red}{\rule{1.5pt}{0.9pt}}%
\textcolor{white}{\rule{1.1pt}{0.9pt}}%
\textcolor{red}{\rule{1.5pt}{0.9pt}}%
\textcolor{white}{\rule{1.1pt}{0.9pt}}%
\textcolor{red}{\rule{1.5pt}{0.9pt}}%
\textcolor{white}{\rule{1.1pt}{0.9pt}}%
: assortative networks
\caption{(Color online) Time evolution of the global state $h^Ts$ toward the
  asymptotic state $h^Ts_\infty$ during social diffusion on
  networks. The absolute difference between $h^Ts$ and $h^Ts_\infty$
  is plotted over time. Black solid, blue dashed, and pink dotted curves show results with
  non-assortative, disassortative and assortative topologies,
  respectively. Each curve shows the mean of 100 independent numerical
  simulation runs, with shaded areas representing standard deviations.
  Each simulation was conducted using Eq.~(\ref{socialdiffusion}),
  starting with a randomly generated initial condition with $n
  =1,000$, $c = 1$, node state range $[-1, 1]$ and link weight range
  $[0, 10]$. Numerical integration was conducted for $t = 0 \sim 10$
  using a simple Euler forward method with step size $\delta t =
  0.01$. (a): Results on random networks. (b): Results on scale-free
  networks. Network topologies were generated first as an unweighted
  network by using the Erd\H{o}s-R\'enyi (for random) or
  Barab\'asi-Albert (for scale-free) network generation
  algorithm. Disconnected networks were not used for the
  experiment. Self-loops were randomly added to nodes with 1\%
  probability. Once the topology was generated, a random link weight
  was assigned to each link. For assortative/disassortative networks,
  a revised Xulvi-Brunet \& Sokolov algorithm
  \cite{xulvibrunet,barabasibook} was additionally used to tune their
  strength assortativity. Assortative (or disassortative) network
  topologies were created by applying to a randomly generated non-assortative
  network 30,000 times of possible link rewirings that would enhance
  degree assortativity (or disassortativity) while preserving node
  degrees. Any link rewiring that would disconnect the network or
  create a multi-link was forbidden. Self-loops were allowed in the
  above operations. Every time a new link was created between
  originally disconnected nodes, a random link weight was assigned to
  the new link.}
\label{speed}
\end{figure*}

We conducted numerical simulations of social diffusion on random and
scale-free networks with non-assortative, disassortative and assortative
topologies. The results are shown in Fig.~\ref{speed}. While the
simulated networks did not have correlation exponents as extreme as
$\mu = \pm 1$ assumed in the analysis above, the simulation results
agree with the analytical predictions, especially during the initial
time period when the networks tend to show a significant global drift
on non-assortative and disassortative networks.

\section{Adaptive Link Weight Adjustment to Promote Upward Global Drift}

The analytical results presented above suggest that, if the local
averages of people's self/neighbor strength ratios are positively
correlated with their states and if the network's strength assortativity
is negative, then an upward drift of the global state will occur
quickly.

Here, we propose an adaptive network \cite{gross2009adaptive} method
of preferential link weight adjustment based on node states and
strengths in order to promote upward global drift while social diffusion
is ongoing. Specifically, we let each pair of nodes $i$ and $j$
dynamically change their connection weight $a_{ij}$ according to the
following dynamical equation,
\begin{equation}
\frac{da_{ij}}{dt} = a_{ij} \left( \alpha \frac{s_i + s_j - 2 \langle s
  \rangle}{2 \sigma_s} - \beta \frac{(k_i - \langle k \rangle) (k_j
- \langle k \rangle)}{\sigma_k^2} \right) ,
\end{equation}
where $\sigma_s$ and $\sigma_k$ are the standard deviations of node
states and link weights, respectively. The first term inside the
parentheses represents the change of link weights to induce positive
correlation between node states and strengths (which naturally promotes
positive correlation between node states and local averages of their
self/neighbor strength ratios), while the second one is to induce
negative strength assortativity.

\begin{figure*}
\centering
\begin{tabular}{ccc}
\includegraphics[width=.49\textwidth]{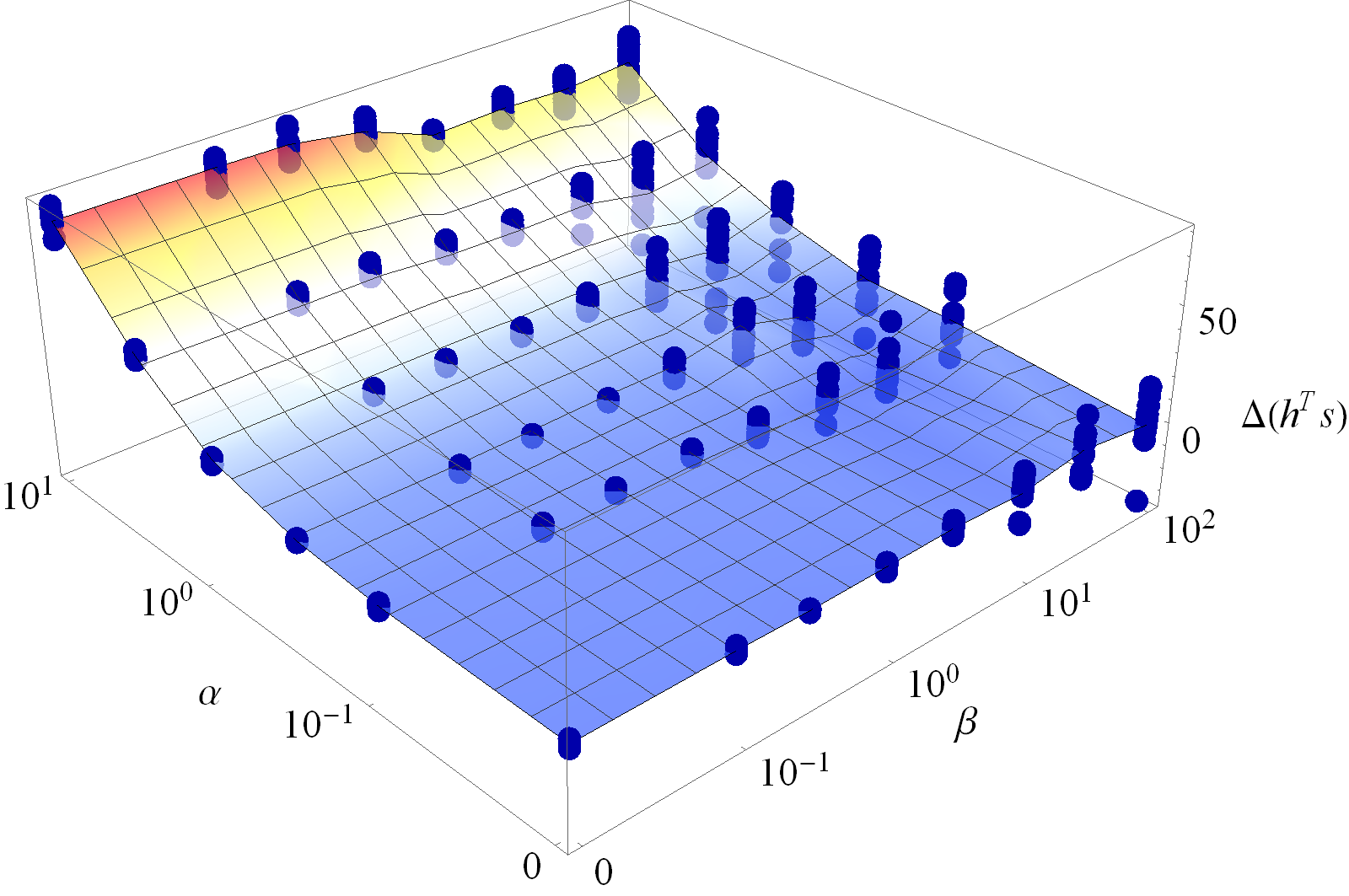} & ~ &
\includegraphics[width=.49\textwidth]{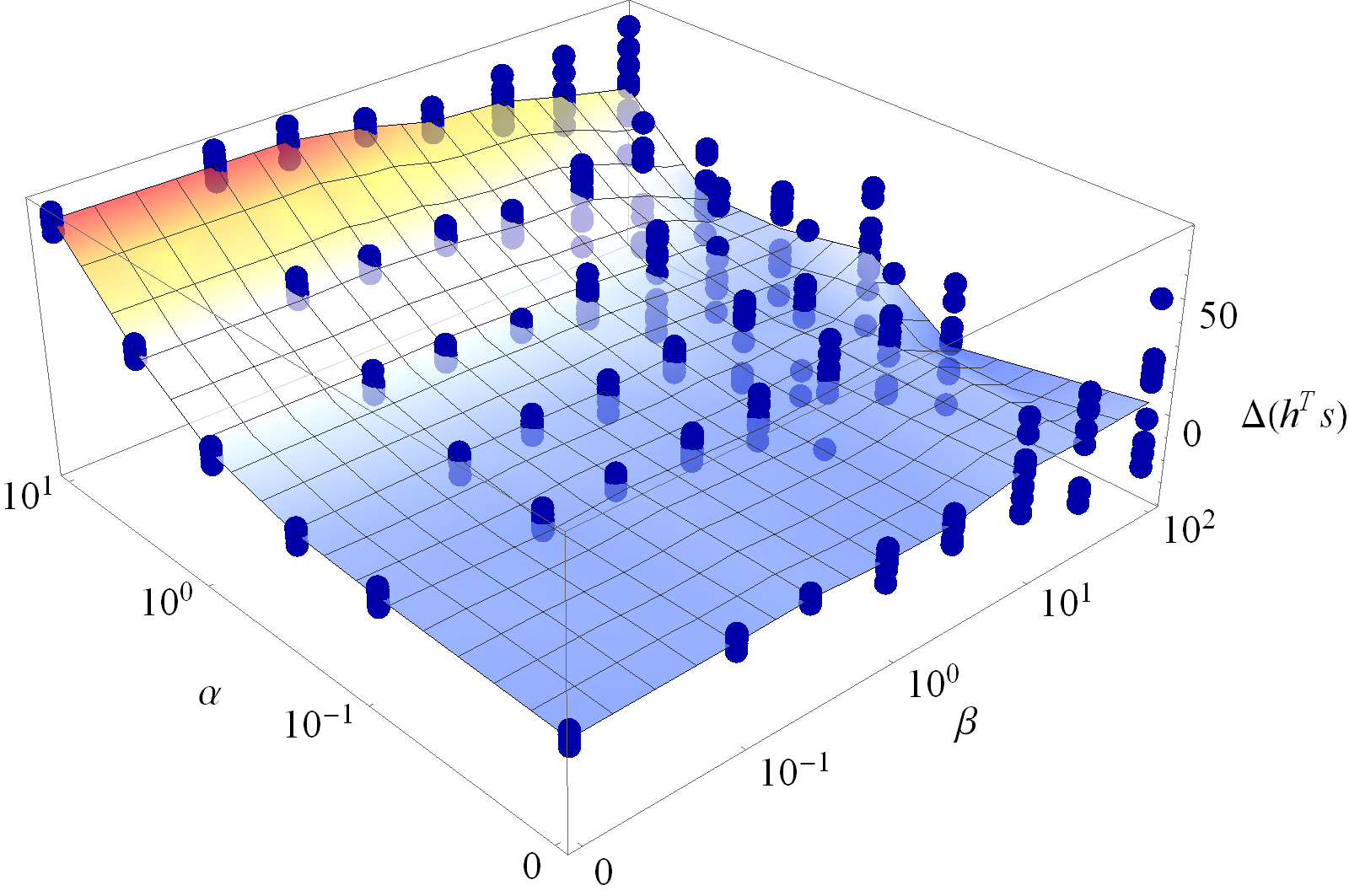} \\
(a) (initially) random & ~ & (b) (initially) scale-free
\end{tabular}
\caption{(Color online) Summary of systematic simulations of social diffusion with
  adaptive link weight adjustment with logarithmically varied $\alpha$
  and $\beta$ (cases with $\alpha = 0$ or $\beta=0$ are also added to
  the lower ends of the axes). The difference in the global state
  between before and after each simulation ($\Delta h^Ts$) is plotted
  over the $\alpha$-$\beta$ parameter space. Ten independent
  simulation runs were conducted for each parameter setting. Each dot
  represents one simulation run, while the surface shows the trend of
  the average. The initial network topology was generated using the
  Erd\H{o}s-R\'enyi (a) or Barab\'asi-Albert (b) network generation
  algorithm, each with 200 nodes and 20\% connection density. Node
  states and link weights were initially random in the same way as in
  the previous experiment (Fig.~\ref{speed}). $c = 1$. Numerical
  integration was conducted for $t = 0 \sim 1$ using a simple Euler
  forward method with step size $\delta t = 0.01$.}
\label{adaptive-results}
\end{figure*}

We examined the effectiveness of this method for promotion of upward
global drift through numerical simulations starting with an initially
random or scale-free network topology with no prior strength-state
correlation or strength assortativity. Systematic simulations were
conducted with $\alpha$ and $\beta$ varied logarithmically. Results
are shown in Fig.~\ref{adaptive-results}. It was found that the
induction of positive strength-state correlation ($\alpha > 0$)
significantly promoted upward global drift. In the meantime, the
induction of negative strength assortativity ($\beta > 0$) increased the
variability of outcomes but did not have a substantial influence on
the direction of the global drift. The highest upward global drift was
achieved when $\alpha$ was large and $\beta$ was small, i.e., when
adaptive link weight adjustment was used merely for inducing positive
strength-state correlation but not for inducing negative strength
assortativity.

\section{Conclusions}

In this paper, we studied the drift of the global state of a network
caused by social diffusion. We showed that the inner product of strength
and state vectors is a conserved quantity in social diffusion, which
plays an essential role in determining the direction and asymptotic
behavior of the global drift. We also showed both analytically and
numerically that the strength assortativity of network topology has a
negative effect on the speed of homogenization. We numerically
demonstrated that manipulation of strength-state correlation via
adaptive link weight adjustment effectively promoted upward drift of
the global state.

This study has illustrated the possibility that social diffusion may
be exploited for individual and collective improvement in real-world
social networks. Mechanisms like the adaptive link weight adjustment
used in the simulations above may be utilized in practice to, for
example, help spread desirable behaviors and/or suppress undesirable
behaviors among youths.

One particularly interesting application area of social diffusion is
education. We have studied possible social diffusion of academic
success in high school students' social network
\cite{blansky2013spread}. This naturally led educators to ask how one
could utilize such diffusion dynamics to improve the students' success
at a whole school level. Na\"ive mixing of students may not be a good
strategy due to bidirectional effects of social diffusion. Our results
suggest that carefully inducing correlation between strengths (amount of
social contacts) and states (academic achievements) would be a
promising, implementable practice at school by, e.g., allowing
higher-achieving students to participate in more extracurricular
activities.

The work presented in this paper still has several limitations that
will require further study. We have so far considered simple symmetric
networks only, but social diffusion can take place on asymmetric,
weighted social networks as well. Our analysis of
assortative/disassortative network cases used unrealistic extreme
assumptions ($\mu = \pm 1$) and ignored topological constraints such
as structural cutoffs that are inevitable for assortative cases
\cite{barabasibook}. Further exploration in these fronts will be
necessary to obtain more generalizable understanding of the social
diffusion dynamics on more complex real-world networks.

\section*{Acknowledgments}

We thank Albert-L\'aszl\'o Barab\'asi for valuable comments. This
material is based upon work supported by the US National Science
Foundation under Grant No.~1027752. R.S.\ acknowledges support from
the James S.\ McDonnell Foundation.

\bibliographystyle{apsrev4-1}
\bibliography{sayama-sinatra}

\end{document}